\documentclass[useAMS,usenatbib, epsfig]{mn2e}
\include{psfig}

\title[On the bispectrum of COBE and WMAP]{On the bispectrum of COBE and WMAP}
\author[Jo\~{a}o Magueijo and Jo\~{a}o Medeiros]{
Jo\~{a}o Magueijo and Jo\~{a}o Medeiros
\thanks{E-mail:j.magueijo@ic.ac.uk; joao.medeiros@ic.ac.uk }\\\\
Theoretical Physics, Imperial College, Prince Consort Road, 
London SW7 2BZ, UK
}
\begin{document}

\date{}

\maketitle

\pagerange{\pageref{firstpage}--\pageref{lastpage}} \pubyear{2003}

\label{firstpage}

\begin{abstract}
The COBE-DMR 4-year maps displayed a strong non-Gaussian signal
in the ``inter-scale'' components of the bispectrum: their observed 
values did not display the scatter expected from Gaussian maps. 
We re-examine this and other suggested non-Gaussian features in the
light of WMAP. We find that they all disappear. Given that it 
was proved that COBE-DMR high noise levels and documented systematics
could at most {\it dilute} the observed non-Gaussian features,   
we conclude that this dataset must have contained non-negligible 
undocumented systematic errors. It turns out that the culprit 
is a combination of QuadCube pixelization and data collected during
the ``eclipse season''.
\end{abstract}


\section{Introduction}\label{introduction}
The possibility of non-Gaussianity in the COBE-DMR 4 year maps
led to a rather protracted  story. Using ``single-$\ell$''
bispectrum analysis,
\citet{fmg} found strong evidence for non-Gaussianity
in the anisotropy of the Cosmic Microwave Background (CMB)
temperature.   This detection was followed by similar claims
by \citet{nfs98} and \citet{pvl98}, and caused considerable consternation
among theorists (see \cite{nvs} for a discussion). Further work,
however, showed that claims other than those based on the bispectrum
could not be reproduced \citep{muk}.
The bispectrum claims were confirmed
by~\cite{bt99}.

Later~\cite{banday} cast serious doubts upon the cosmological
origin of the observed signal. A systematic was identified which
removed the observed ``single-$\ell$'' bispectrum  signal -- the so-called
eclipse season data, which should never have been used. However,
when an extension to ``inter-$\ell$'' bispectrum components was
sought, a new non-Gaussian signal was found for inter-$\ell$ separations
of $\Delta l=1$~\citep{joao}. Specifically, their observed values 
were found to concentrate uncannily close to zero instead of displaying 
the scatter expected from Gaussian maps.

This signal could not be blamed 
on any systematic effects studied by~\cite{banday} or otherwise \citep{joao}. 
It was also proved that the instrument high noise levels could at most
dilute the observed signal, in spite of the noise correlations and
anisotropy \citep{joao}. It was also found that the observed signal
did not extend to higher inter-$\ell$ separations $\Delta l>1$ 
\citep{haav}.

What shall we make of these claims in the light of the recent 
observations \citep{wmap} by the Wilkinson Microwave Anisotropy Probe (WMAP)? 
In this paper we show that they don't survive the 
new data, which displays consistency with Gaussianity on the 
angular scales probed by COBE. This might seem at first suprising, 
given that the WMAP data is considerably less noisy than the COBE-DMR
dataset, and that noise can at most hide a non-Gaussian signal. However 
one should never forget the issue of systematics. Even though documented
COBE-DMR systematics were shown not to correlate with the observed 
non-Gaussian signal, the new data allows us to cast a new look at the
problem. We find that a highly non-subtle combination of QuadCube
pixelization systematics and the ``eclipse data''  are  to
be blamed for the observed effect.

This, we believe, closes the story. The moral is clear: care must
be exercised regarding similar claims currently being made with
WMAP.

\section{The bispectrum}
We start by reviewing some results and
definitions pertaining to the bispectrum. Given a full-sky map,
$\frac{\Delta T}{T}({\bf n})$, this may be expanded into Spherical
Harmonic functions:
\begin{eqnarray}
\frac{\Delta T}{T}({\bf n})=\sum_{\ell m}a_{\ell m}Y_{\ell m}({\bf n})
\label{almdef}
\end{eqnarray}
The coefficients $a_{\ell m}$ may then be combined into
rotationally invariant multilinear forms (see \cite{santa}
for a possible algorithm). The most general
cubic invariant is the bispectrum,  and is given by
\begin{eqnarray}
{\hat B}_{\ell_1\ell_2\ell_3}&=&\frac{\left (
\begin{array}{ccc} \ell_1
 & \ell_2 & \ell_3 \\ 0 & 0 & 0
\end{array} \right )^{-1}}
{(2\ell_1+1)^{\frac{1}{2}}
(2\ell_2+1)^{\frac{1}{2}}(2\ell_3+1)^{\frac{1}{2}}}\times \\
&&\sum_{m_1m_2m_3}\left
( \begin{array}{ccc} \ell_1 & \ell_2 & \ell_3 \\ m_1 & m_2 & m_3
\end{array} \right ) a_{\ell_1 m_1}a_{\ell_2 m_2} a_{\ell_3 m_3}\nonumber
\end{eqnarray}
where the $(\ldots)$ is the Wigner $3J$ symbol.
The proportionality constant is usually chosen 
in order to enforce a roughly constant cosmic variance. 
In \cite{fmg} the choice was made  $\ell_1=\ell_2=\ell_3$, leading to
the ``single-$\ell$'' bispectrum ${\hat B_\ell}=B_{\ell\, \ell \,\ell}$. 
Other bispectrum components 
are sensitive to correlations between different scales.
Selection rules require that $\ell_1+\ell_2+\ell_3$ be even.
The simplest chain of correlators is therefore
${\hat A_\ell}=B_{\ell-1\, \ell \,\ell+1}$ -- the ``inter-$\ell$
bispectrum -- and this was studied in~\cite{joao}.
Other components, involving more distant multipoles, 
may be considered \citep{haav}
but they are very likely to be dominated by noise; it is
natural to assume that possible non-Gaussian inter-scale
correlations decay with $\ell$ separation.

We shall therefore consider ratios
\begin{equation}\label{i3}
I^3_\ell={ {\hat B}_{\ell}
\over ({\hat C}_{\ell})^{3/2}}
 \label{defI}
\end{equation}
and
\begin{eqnarray}\label{j3}
J^3_\ell &=& { {\hat A}_{\ell}
\over ({\hat C}_{\ell-1})^{1/2}({\hat C}_{\ell})^{1/2}
({\hat C}_{\ell+1})^{1/2}}
 \label{defJ}
\end{eqnarray}
where ${\hat C}_\ell=\frac{1}{2\ell+1}\sum_m|a_{\ell m}|^2$.
These quantities are dimensionless, and therefore less
dependent upon the power spectrum. They are also invariant under rotations
and parity.

The theoretical importance of the bispectrum as a non-Gaussian
qualifier has been  recognized in a number of publications
(\cite{luo94}, \cite{peebles1},
\cite{sg98}, \cite{gs98}, \cite{wang}).
\cite{kog96a} measured the pseudocollapsed
and equilateral three point function of the DMR four year data.
The bispectrum may be
regarded as the Fourier space counterpart of the three point function.

\section{Bispectrum analysis of WMAP data }

The WMAP mission \citep{wmap} was designed to make full sky CMB maps with
unprecedented accuracy. There are ten differencing assemblies (DAs)
in total, four in the W band at 94 GHz, two V band at 61 GHz, two
Q band at 41 GHz, one Ka band at 33 GHz, and one K band at 23 GHz.
The K and Ka bands are dominated by galactic emission and
therefore neglected for cosmological analysis. The maps are made
using the HEALPix
 \footnote{ The HEALPix website is http://www.eso.org/science/healpix} format with nside=512 \citep{healp,healp1}.
The total number of pixels in each map is $12
\times nside^2 = 3, 145, 728$. We use the coadded sum map of the
Q, V and W maps,
\begin{equation}\label{coadd}
 T =
\frac{\sum^{10}_{i=3} T_{i} /
\sigma^{2}_{0,i}}{\sum^{10}_{i=3} 1 / 
\sigma^{2}_{0,i}}
\end{equation}
where $T_{i}$ is the sky map for the DA $i$ with the foreground
galactic signal subtracted, and $\sigma^{2}_{0,i}$ is the noise
per observation for DA i, whose values are given by \cite{wmap}. We use the publicly available 'foreground cleaned' maps,
where the Galactic foreground signal, consisting of synchrotron,
free-free, and dust emission, was removed using the 3-band,
5-parameter template fitting method described in \cite{Bennett:2003ca}.
We then use the Kp0 mask to cut the Galactic plane emission and the known
point sources (\cite{Bennett:2003ca} ), retaining 76.8\% of the
sky.

The monopole and dipole are removed and we perform a harmonic
analysis of the map obtaining the $a_{\ell m}$ up to $l=20$. This is
performed using the FORTRAN utility ANAFAST, available in the
HEALPix package. We then evaluate (\ref{j3}) and the 
$J^{3}_{l}$ obtained for the WMAP data are
compared to the distributions $P(J^{3}_{l})$ obtained from
Gaussian simulations subject to the appropriate beam, galactic mask
and noise. These simulations take as input the LCDM power-law primordial 
power spectrum fit to the
WMAP, CBI and ACBAR data (\cite{Bennett:2003ca}). The random fields were generated using the utility
SYNFAST of the HEALPix package. The distributions so obtained
do not vary significantly from those obtained previously for COBE.

\section{Results and discussion}

\begin{figure}
\centerline{\psfig{file=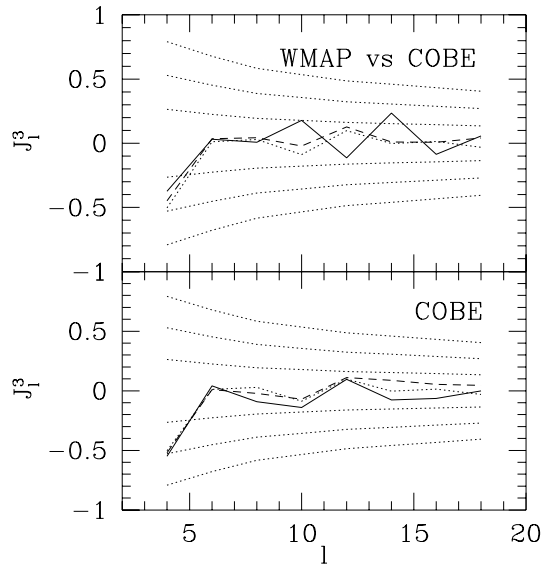,width=8cm}}
 \caption{The COBE inter-$\ell$ non-Gaussian signal (bottom), using 
three different data renditions: HEALPix (solid line) and Quad-Cube
ecliptic (dashed) and galactic (dotted). In the top pannel we have
re-reproduced the COBE galactic Quad-cube data (dotted), and superposed
a COBE map in the same pixelization but without the eclipse data
(dashed), and the WMAP results (solid). The continuum dotted lines
represent 1, 2 and 3 sigma contours for the bispectrum expected from
Gaussian maps.}
\label{fig1}
\end{figure}
In Fig.~\ref{fig1} we plot the inter-$\ell$ bispectrum of several
renditions of the COBE data, and of WMAP. The original COBE-DMR 
non-Gaussian inter-$\ell$ signal \citep{joao} was found using 
QuadCube pixelization. Possible deficiencies of this scheme
were evaluated by aligning the pixelization system with galactic 
and ecliptic coordinates. The
$J^{3}_{l}$ in both frames are plotted in the bottom panel of
Fig.~\ref{fig1}. The dotted contours represent 1, 2 and 3 sigma 
lines for the bispectrum arising from Gaussian realizations. 

One would expect to see one in three coefficients lying outside 
the 1-sigma contour. Instead, for COBE-DMR maps we see a very close alignment
with the peak of the distribution. This gives a reduced chi squared
$X^2=0.14$ and $X^2=0.22$ for data in galactic and ecliptic 
pixelization, respectively. Computing the distributions
$P(X^2)$ leads to $P(X^2>0.14)= 0.998$ 
(and $P(X^2>0.22)=0.985$) for maps in galactic (ecliptic) pixelization. 
These are the confidence levels for rejecting Gaussianity
on the grounds of the COBE bispectrum.

As can be seen from the upper panel in Fig.~\ref{fig1} the WMAP
bispectrum does not have such an obvious lack of scatter. Indeed it
leads to $X^2=0.59$, consistent with Gaussianity.

What can be the origin of this discrepancy? It is at once clear 
that we are not comparing like with like. The WMAP project used the 
HEALPix pixelization scheme, and the tests made for the impact
of QuadCube upon the COBE map may not be conclusive. For example
there may be an isotropic systematic effect, present no matter how
one orients the coordinate system. To address this in Fig.~\ref{fig1} 
we plotted (solid line in bottom panel) the $J^{3}_{l}$ from
COBE 4 year maps rendered in HEALPix. 
We found $X^2=0.26$, reducing the confidence level to 96\%.
Hence the COBE/WMAP discrepancy can partly be blamed
on a poor pixelization scheme, but this is not enough: even in the 
HEALPix rendition COBE is much more non-Gaussian than WMAP. 
\begin{figure}
\centerline{\psfig{file=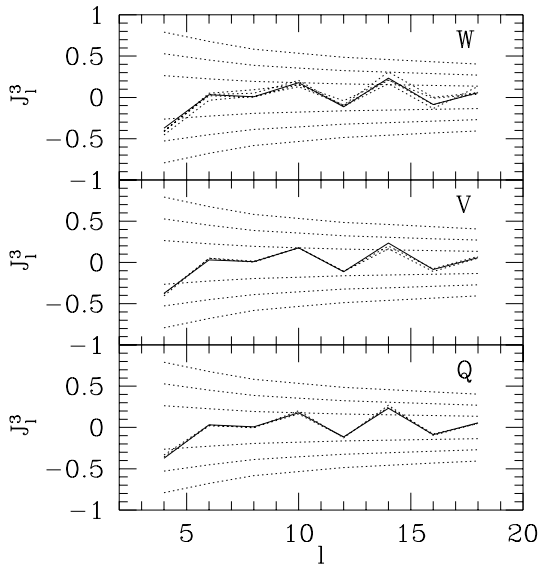,width=8cm}}
 \caption{The inter-$\ell$ spectrum for the various WMAP Q, V,
and W bands (dotted lines). The solid line on all pannels is the 
spectrum for the co-added maps.}
\label{fig2}
\end{figure}

Could we have missed the non-Gaussian signal in WMAP by looking 
at the wrong combination of frequencies? Equation (\ref{coadd})
favours the Q band, which is the most contaminated by galactic emission. 
Galactic emissions have been proved to degrade non-Gaussian large
angle signals \citep{joao}. However, as can be
seen in Fig.~\ref{fig2} there is a remarkable consistency between
the bispectrum observed in all WMAP channels, a tribute to the
very high signal to noise, but also to the lack of galactic 
contamination. By way of contrast the COBE inter-$\ell$ non-Gaussian
signal came mainly from the 53 GHz  channel,  which was also the least 
noisy channel.

\begin{figure}
\centerline{\psfig{file=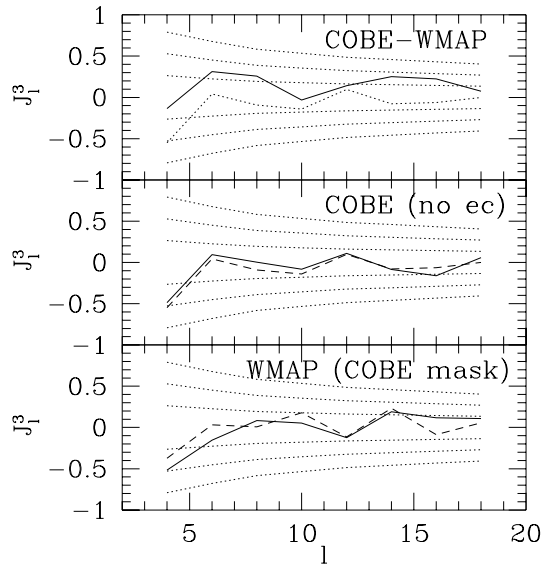,width=8cm}}
\caption{The bottom pannel shows the $J^3_\ell$
for WMAP subject to the COBE mask (solid
line) and to the  Kp0 mask (dotted). The middle
panel shows the $J^3_\ell$ for the COBE maps in 
HEALPix with (dashed) and without (solid) the eclipse
data. The top panel (solid line)
shows the inter-$\ell$ spectrum of a map obtained by subtracting 
WMAP to COBE (both with the COBE mask). We have added the COBE (HEALPix) bispectrum
for reference. }
\label{fig3}
\end{figure}

A related possible source of discrepancy is the galactic
mask. The Kp0 mask is significantly smaller than the extended
galactic cut used in COBE \citep{banday97}. Could the extra 
regions near the galactic plane hide a non-Gaussian signal? 
As the bottom pannel in Fig.~\ref{fig3} shows this is not the case.
Subjecting the WMAP data to the extended cut used by the COBE team
in fact increases the WMAP chi squared to $X^2=0.64$. 

It would therefore appear that nothing has been missed, and that the 
WMAP bispectrum on ``COBE'' large angular scales is indeed
consistent with Gaussianity. 
Given that the higher noise levels in the COBE maps
can at most dilute a non-Gaussian signal (a fact proved
in \cite{joao} even after taking noise correlations into
account), we may conclude that a systematic 
error is behind the COBE non-Gaussian signal. 

Nevertheless, identifying
the culprit is far from obvious: no documented 
COBE systematic mimics the
observed inter-$\ell$ signal. The effects of the eclipse data
\citep{banday} on the $J^{3}_{l}$,
for example, are assessed in the top panel of Fig.~\ref{fig1}.
They give $X^2=0.18$, leading a confidence level
for rejecting Gaussianity of $99.2\%$. 
It turns out, however, that if we reject data collected
during the eclipse season  {\it and} construct a 
DMR 4-year map in HEALPix the puzzle is solved. This is
shown in the middle panel in Fig.~\ref{fig3}, where we have
plotted the $J^{3}_{l}$ inferred from such (co-added) maps.
Although the difference might look subtle, there is quite
a substantial difference around $\ell=16$. In fact the
reduced chi squared in now $X^2=0.42$, consistent with
Gaussianity.

Curiously the single-$\ell$ non-Gaussian signal found by  \cite{fmg} 
results mainly from $I^3_{\ell}$ at $\ell=16$. This stops being a 
severe deviant once the eclipse data is excluded. It would now appear 
that the same
happens for the $J^3_\ell$, but only after a better pixelization
scheme is introduced.

\begin{figure}
\centerline{\psfig{file=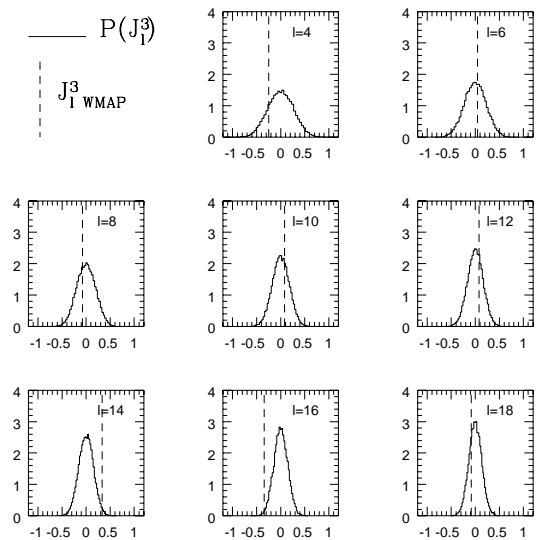,width=8cm}}
\caption{The vertical thick dashed line is the value of the observed $J^{3}_{l}$. The solid line histogram is the pdf of the  $J^{3}_{l}$ obtained from 4020 Gaussian simulations of the sky with noise and the Kp0 mask. The dashed line histogram superimposed is the same pdf obtained with the COBE's extended galactic cut and DMR noise. }
\label{fig4}
\end{figure}

\section{Conclusions}
The conclusion is that a combination of pixelization effects
and the ``eclipse'' systematic is behind the inter-$\ell$
non-Gaussian effect previously reported for COBE maps. Curiously
if we take a difference map (COBE minus WMAP) the anomalous alignment of 
the $J^{3}_{l}$ is not present (see top panel of Fig.~\ref{fig3}).
On the scales we are considering such maps picture COBE noise, plus COBE
systematics, minus WMAP systematics (assumed to be small). Thus
one would expect the COBE $J^{3}_{l}$ signal to be enhanced in the
difference map. The fact that it is not results from the non-linearity
of the statistic being used, plus the subtle interplay of signal
and systematic via the QuadCube pixelization scheme. 

In this paper we have concentrated on $J^3_\ell$, because the single
scale bispectrum anomalies have been explained long ago. However we 
have checked that no new anomalies on large angular scales emerge 
in the WMAP data. 

We reserve to a future publication a complete study of the bispectrum
of WMAP of smaller angular scales. To our mind the work of \cite{ngwmap}
is just the beginning.

\section*{Acknowledgments}
We would like to thank A.J. Banday for help with this project and for
supplying COBE-DMR maps in HEALPix, and without the eclipse data.
We thank E. Hivon. E. Komatsu, K. Land and A. Jaffe for comments and 
discussion. J. Medeiros would also like to thank the help of Gra\,ca Rocha and P.Ferreira and the  support given by S.Rankin. Some of the results in this paper have been derived using the HEALPix package (\cite{healp}). J. Medeiros also acknowledges the financial support of Fundacao para a Ciencia e Tecnologia (Portugal).

\bsp

\label{lastpage}

\end{document}